\documentclass[aps,preprint]{revtex4}%
\usepackage{amsfonts}
\usepackage{amsmath}
\usepackage{amssymb}
\usepackage{graphicx}%
\providecommand{\U}[1]{\protect\rule{.1in}{.1in}}
\newtheorem{theorem}{Theorem}
\newtheorem{acknowledgement}[theorem]{Acknowledgement}

\begin{document}
\title[Time operator]{A dynamical time operator in Dirac's relativistic quantum mechanics$^{a}$}
\author{M. Bauer}
\affiliation{Instituto de F\'{\i}sica, Universidad Nacional Aut\'{o}noma de M\'{e}xico.
Apdo. Postal 20-364, 01000 M\'{e}xico DF, M\'{e}xico}
\email{bauer@fisica.unam.mx}
\keywords{Relativistic quantum mechanics, dynamical time operator, time-energy
uncertainty, space-time confinement, Zitterbewegung Relativistic quantum
mechanics, dynamical time operator, time-energy uncertainty, space-time
confinement, Zitterbewegung Relativistic quantum mechanics, dynamical time
operator, time-energy uncertainty, space-time confinement, Zitterbewegung
Relativistic quantum mechanics, dynamical time operator, time-energy
uncertainty, space-time confinement, Zitterbewegung}
\pacs{PACS number}

\begin{abstract}
A self-adjoint dynamical time operator is introduced in Dirac's relativistic
formulation of quantum mechanics and shown to satisfy a commutation relation
with the Hamiltonian analogous to that of the position and momentum operators.
The ensuing time-energy uncertainty relation involves the uncertainty in the
instant of time when the wave packet passes a particular spatial position and
the energy uncertainty associated with the wave packet at the same time, as
envisaged originally by Bohr. The instantaneous rate of change of the position
expectation value with respect to the simultaneous expectation value of the
dynamical time operator is shown to be the phase velocity, in agreement with
de Broglie's hypothesis of a particle associated wave whose phase velocity is
larger than c. Thus, these two elements of the original basis and
interpretation of quantum mechanics are integrated into its formal
mathematical structure. Pauli's objection is shown to be resolved or
circumvented. Possible relevance to current developments in interference in
time, in Zitterbewegung like effects in spintronics, grapheme and
superconducting systems and in cosmology is noted

\textbf{Keywords:} Relativistic quantum mechanics, dynamical time operator,
time-energy uncertainty, space-time confinement, Zitterbewegung Relativistic
quantum mechanics, dynamical time operator, time-energy uncertainty,
space-time confinement, Zitterbewegung

\end{abstract}
\maketitle

\section{Introduction}

\textbf{\textquotedblleft One must be prepared to follow up the consequences
of theory, and feel that one just has to accept the consequences no matter
where they lead\textquotedblright\ }P. A. M. Dirac$^{b}$.

In the formulation of quantum mechanics (QM) time appears as a parameter, not
as a dynamical variable. It is a c-number, following Dirac's
designation\cite{Dirac}. Thus QM fails to treat time and space coordinates on
the almost equal footing accorded by the special theory of relativity, as it
does with momentum and energy. As an explanation, it has interestingly been
asserted that \textquotedblleft time always arises in quantum mechanics as an
externally defined classical parameter from the interaction with a classical
environment\textquotedblright. Indeed it is shown that the time independent
Schr%
\"{}%
odinger or Dirac equations describing system and environment, give rise to the
corresponding time dependent equations in a disentangling approximation where
\textquotedblleft the motion of the environment provides a time derivative
which monitors the development of the quantum system\textquotedblright.
Consequently \textquotedblleft time enters quantum mechanics only when an
external force on the quantum system is considered
classically\textquotedblright\cite{Briggs1},\cite{Briggs2}.

The above then still leaves open the question of a \textquotedblleft time
operator\textquotedblright\ $T$ satisfying a commutation relation%

\begin{equation}
\lbrack T,H]=i\hbar,\qquad
\end{equation}

as the one satisfied by the position and momentum operators. Its existence has
had to deal with the fundamental objection pointed by Pauli, that the finite
lower bound of the energy spectrum precludes mathematically the existence of a
self adjoint operator (\textquotedblleft i.e., as a function of q and
p\textquotedblright) canonically conjugate to the Hamiltonian.\cite{Pauli}

As a consequence, the role of time in quantum mechanics as well as the
existence of time operators and the diverse formulations and interpretations
of a time-energy uncertainty relation have been the subject of extensive
investigations since\cite{Muga}. To be noted in particular is the proposal of
Aharonov and Bohm\cite{Aharonov}, in the framework of the non relativistic
Schr\"{o}dinger equation for the free particle Hamiltonian $H=p^{2}/2m$ , of a
maximally symmetric time operator $(1/2m)\{(1/p)x+x(1/p)\} $ associated with a
\textquotedblleft time-of-arrival\textquotedblright\ concept and needing the
acceptance of Positive Operator Valued Measures (POVM), an extension of the
standard von Neumann definition of \textquotedblleft
observables\textquotedblright\cite{Aharonov}.

Also to be noted is that, within the time quantities considered, such as
parametric (clock) time, tunneling times, decay times, dwell times, delay
times, arrival times or jump times, one finds both instantaneous values and
intervals. To quote the introduction in Ref. 5, \textquotedblleft In fact, the
standard recipe to link the observables and the formalism does not seem to
apply, at least in an obvious manner, to time observables\textquotedblright.

In previous work\cite{Bauer} however, it was shown that Pauli's objection is
overcome formally by enlarging the Hilbert space, obviously by continuing the
energy spectrum to negative energies but also, equivalently, by introducing a
spin-like quantum number so as to associate two states to each positive energy
value. In both of these enlarged spaces, a unitary energy displacement
operator can be introduced whose generator is a \textquotedblleft time
operator\textquotedblright\ that satisfies the above commutation relation. Its
expectation value is equal to plus or minus the evolution parameter $t$,
corresponding to the negative energy extension and to the positive energy
extension, respectively.

In the present work, the relativistic free particle Dirac Hamiltonian is the
starting point, instead of the non relativistic one. It suggests a particular
form for a \textquotedblleft dynamical time operator\textquotedblright, to be
denoted $T(t)$, i.e., dependent on the parametric time and introducing a new
constant that would be a characteristic of the system. In Section II, the
corresponding $[T,H]$ commutator is evaluated and the time evolution of the
expectation value is derived from the dynamical postulate of QM. In Section
III, the Heisenberg picture is used to establish its dependence on the
parametric time. On the basis of its eigenvalues and eigenvectors derived in
Appendix B, Section IV contains a clarification of the corresponding
time-energy uncertainty relation. Section V addresses Pauli's objection. The
behavior with respect to the discrete symmetries, parity, time reversal and
charge conjugation, is considered next (Section VI), where the role of the new
constant is clarified. Its value is suggested by the interpretation of the
ensuing Zitterbewegung behavior, analysed in Section VII. Finally, Section
VIII contains conclusions and possible relevance to current research areas.

\section{A \textquotedblleft time operator\textquotedblright\ in RQM}

In analogy with the free particle Dirac Hamiltonian\cite{Dirac},\cite{Pauli}%
,\cite{Capri},\cite{Thaler}%
\begin{equation}
H=c\boldsymbol{\alpha\cdot p}+\beta m_{0}c^{2}%
\end{equation}

where $\boldsymbol{\alpha}=(\alpha_{x},\alpha_{y},\alpha_{z})$ \ and $\beta$
are the Dirac matrices, the following expression for a self-adjoint time
operator is considered:%

\begin{equation}
T=\boldsymbol{\alpha\cdot r}/c+\beta\tau_{0},
\end{equation}

where $\tau_{0}$ is a real constant with dimensions of time, and in principle
a property of the system. From the commutation relation between position and
momentum $[x_{i},p_{j}]=i\hbar\delta_{ij}$and the properties of the Dirac
matrices, it follows:%
\begin{equation}
\lbrack T,H]=i\hbar\{3I+4\boldsymbol{s\cdot l}/\hbar^{2}\}+2\beta\{\tau
_{0}(c\boldsymbol{\alpha\cdot p})-m_{0}c^{2}(\boldsymbol{\alpha\cdot r}/c)\}
\end{equation}

where $I$, $\boldsymbol{s}$ and $\boldsymbol{l}$ are the identity, spin and
orbital angular momentum operators, respectively. Introducing the
\textquotedblleft spin-orbit\textquotedblright\ operator $K=\beta
(2\boldsymbol{s\cdot l}/\hbar^{2}+1)$ , that commutes with $\boldsymbol{j=l+s}%
$ and $H$ , and is therefore a constant of motion\cite{Capri},\cite{Thaler},
one has%
\begin{equation}
\lbrack T,H]=i\hbar\{I+2\beta K\}+2\beta\{\tau_{0}(c\boldsymbol{\alpha\cdot
p})-m_{0}c^{2}(\boldsymbol{\alpha\cdot r}/c)\}
\end{equation}

or, equivalently%
\begin{equation}
\lbrack T,H]=i\hbar\{I+2\beta K\}+2\beta\{\tau_{0}H-m_{0}c^{2}T\}.
\end{equation}

As such, this commutator is not entirely analogous to the position momentum
commutation relation, but does contain a constant term whose expectation value
is state independent, namely $i\hbar I.$

>From the dynamical postulate of QM, the time evolution of the expectation
value of the time operator is then given by:%
\begin{equation}
d/dt<T>=(1/i\hbar)<[T,H]>=<\{I+2\beta K\}+2\beta\{\tau_{0}H-m_{0}c^{2}T\}>.
\end{equation}

In the presence of a potential$V(r)$ that depends only on position and thus
commutes with the time operator, Eq. (5) is still valid, whereas the form (6)
requires to substitute $H-V(r)$ for $H$ in the last term (Appendix C examines
the case of an electromagnetic field).

As a consequence of the other terms, the time evolution of the time operator
includes the peculiar (Zitterbewegung) time dependence of the position
operator in RQM, as can be seen more clearly in the Heisenberg picture.

\section{ The free particle case in the Heisenberg picture}

In the case of a free particle, where $\boldsymbol{p}$ and $H$ are constants
of motion, one has in the Heisenberg picture\cite{Dirac},\cite{Pauli}%
,\cite{Capri},\cite{Thaler}%
\begin{equation}
\boldsymbol{\alpha}(t)=c\boldsymbol{p}/H+\{\boldsymbol{\alpha}%
(0)-c\boldsymbol{p}/H\}exp(-2iHt/\hbar),\qquad
\end{equation}

\begin{equation}
\beta(t)=m_{0}c^{2}/H+\{\beta(0)-m_{0}c^{2}/H\}exp(-2iHt/\hbar)
\end{equation}

and%
\begin{equation}
\boldsymbol{r}(t)=\boldsymbol{r}(0)+(c^{2}\boldsymbol{p}/H)t+i(c\hbar
/2H)\{\boldsymbol{\alpha}(0)-c\boldsymbol{p}/H\}[exp(-2iHt/\hbar)-1]
\end{equation}

As $(c^{2}p/H)=dE/dp$ represents the group velocity $v_{gp}$, $\boldsymbol{r}%
(t)$ is shown to follow the classical motion (Ehrenfest's theorem), albeit
accompanied by oscillating terms (Zitterbewegung) that nevertheless vanish for
only positive energy or only negative energy wave packets. Using Eqs. (8), (9)
and (10), it follows that (see Appendix B for the full expression):%
\begin{equation}
T(t)=(c\boldsymbol{p}/H).\{\boldsymbol{r}(0)/c^{2}+(c\boldsymbol{p}%
/H)t\}+(m_{0}c^{2}/H)\tau_{0}+oscillatingterms.
\end{equation}

Leaving aside these oscillating terms and introducing explicitly the group
velocity $v_{gp}$ , one has, setting $\boldsymbol{r}(0)=0$ for simplicity:%
\begin{equation}
T(t)=(\boldsymbol{v}_{gp}/c)^{2}t+(m_{0}c^{2}/H)\tau_{0},
\end{equation}

\begin{equation}
\boldsymbol{r}(t)=\boldsymbol{v}_{gp}t.
\end{equation}

It is seen that, although proportional to $t$, in general $T(t)<t.$ Only in
the limit $m_{0}c^{2}=0$, $\boldsymbol{v}_{gp}$ is equal to $\boldsymbol{c}$,
$T(t)=t$ and $\boldsymbol{r}(t)=\boldsymbol{c}t=\boldsymbol{c}T(t)$. Non
relativistic and ultra relativistic limits of the time operator are shown in
Appendix C.

>From Eqs. (10) and (11), it also follows that:%
\begin{equation}
d\boldsymbol{r}(t)/dT(t)=\boldsymbol{v}_{gp}dt/(v_{gp}/c)^{2}%
dt=(\boldsymbol{v}_{gp}/v_{gp})(c/v_{gp})=\boldsymbol{v}_{ph}%
\end{equation}

is the phase velocity $v_{ph}=E/p$, which is collinear with $v_{gp}$, and such
that $v_{ph}v_{gp}=c^{2}$. Consequently $v_{ph}>c$. This agrees with the
property that de Broglie derives for the wave he associates to a material
particle\cite{Broglie}. And indeed it is shown in de Broglie's thesis that the
phase velocity of the proposed wave satisfies $v_{ph}v_{gp}=c^{2}$ , where
$v_{gp}$ is the speed of the \textquotedblleft mobile\textquotedblright\ that
is associated to the transport of energy, i.e., the group velocity of a
superposition of waves with close-by frequencies.

\section{ The time-energy uncertainty relation\cite{Busch}}

A time-energy uncertainty relation can now be derived in the usual way from
the Schawrtz inequality, applied to the uncertainties $(\Delta T)^{2}%
=<T^{2}>-<T>^{2}$, and

$(\Delta H)^{2}=<H^{2}>-<H>^{2}$ of the self-adjoint operators$T$ and $H$ ,
namely:%
\begin{equation}
(\Delta T)^{2}(\Delta H)^{2}\geq(1/4)|<[T,H]>|^{2}\geq(\hbar^{2}/4)|<(I+2\beta
K)>|^{2}%
\end{equation}

As shown in Appendix A, in entire analogy with the eigenvectors and
eigenvalues of the free particle relativistic Dirac Hamiltonian, the
eigenvectors of the time operator are of the form%
\begin{equation}
|\tau>=u_{r}|\boldsymbol{r}>\qquad
\end{equation}

where $|\boldsymbol{r}>$ is the eigenvector of the position operator
$\boldsymbol{r}$ and $u_{r}$ is a four component spinor independent of the
linear momentum $\boldsymbol{p}$. The corresponding doubly degenerate
eigenvalues are%
\begin{equation}
\tau=\pm\ \tau_{r}=\pm\ [(r/c)^{2}+\tau_{0}^{2}]^{1/2}%
\end{equation}

Thus, a wave packet centered about $\tau_{R}=[(R/c)^{2}+\tau_{0}^{2}]^{1/2}$
at a time $t$ and of width $\Delta T$ is actually a wave packet centered at a
point $R$ of width $\Delta r$. Its Fourier transform yields a wave packet in
momentum space of width $\Delta p$ centered at a value $P$, which in turn
represents a wave packet of width $\Delta E$ about $Ep=+[(pc)^{2}+(m_{0}%
c^{2})^{2}]^{1/2}$ . Thus the position momentum uncertainty relation $(\Delta
r)_{t}(\Delta p)_{t}\geq\hbar/2$ derives into a time energy uncertainty
relation $(\Delta T)_{t}(\Delta E)_{t}\geq\hbar/2$ , in agreement with the
commutation relation, Eq.(6). To be emphasized is that the above expectation
values and uncertainties correspond to instantaneous evaluations at time $t$,
in agreement with Bohr's conception, as quoted by Pauli\cite{Pauli}.

The dynamical time operator here proposed is the appropriate one to define the
time of passage or arrival time at a specific point. In contrast, as pointed
earlier, in many of the interpretations of a time-energy uncertainty relation
the $\Delta t$ corresponds to a time interval, not to an instantaneous value
of the uncertainty. Dwell times, tunneling times, i.e., time intervals, should
be expressed as differences of average values of the time operator taken at
two different points of the trajectory, and consequently related to parametric
time differences (Appendix C). Their time uncertainties would need to combine
the instantaneous uncertainties of the end points. On the other hand, the
often quoted uncertainty relation between line width and lifetimes of unstable
states derives from the dynamics generated by the Schr\"{o}dinger
equation\cite{Wigner},\cite{Mello}.

\section{ What about Pauli's objection?}

Considering that the position operator $\boldsymbol{r}$ in momentum space is
the generator of momentum translations, that is,%
\begin{equation}
exp(i\delta\boldsymbol{p\cdot r}/\hbar)|\boldsymbol{p}>=|\boldsymbol{p}%
+\delta\boldsymbol{p}>,\qquad
\end{equation}

the unitary operator%
\begin{equation}
U(\epsilon)=(i\epsilon T/\hbar)=expi\epsilon\{\boldsymbol{\alpha\cdot
r}/c+\beta\tau_{0}\}/\hbar=(1+i\epsilon\boldsymbol{\alpha\cdot r}%
/c\hbar+...)expi\beta(\epsilon\tau_{0}/\hbar)
\end{equation}

where $\epsilon$ is a (positive or negative) infinitesimal energy, generates
both a change in phase by the amount $\beta(\epsilon\tau_{0}/\hbar)$ and a
momentum displacement by the amount $\delta\boldsymbol{p}=(\epsilon
/c)\boldsymbol{\alpha}$ \ in the direction of the instantaneous velocity
$c\boldsymbol{\alpha}=d\boldsymbol{r}/dt$. Averaged over a wave packet, this
can be seen as a \textquotedblleft boost\textquotedblright, that is, a change
to a reference frame where the corresponding energy is shifted by $\delta
E=(\epsilon/c)\boldsymbol{\alpha\cdot v}_{gp}$where $\boldsymbol{v}_{gp}$ is
the group velocity $c^{2}\boldsymbol{p}/H$.

Repeated applications can generate finite displacements over all the momentum
space, and consequently finite energy shifts, without however leaving the
positive (or negative) energy spectrum as the solutions for positive and
negative energy transform separately under proper Lorentz transformations.
Energy goes through a minimum (maximum) as the momentum goes through zero,
remaining either above (or below) the $2m_{0}c^{2}$ energy gap. Both the
positive and negative spectra eigenvalues $\pm\lbrack(pc)^{2}+(m_{0}c^{2}%
)^{2}]^{1/2}$ of the Dirac Hamiltonian are degenerate with respect to
$\boldsymbol{p}$ and $-\boldsymbol{p}$, providing the \textquotedblleft pseudo
spin\textquotedblright\ extension $|E;\sigma>$, with $\sigma=\pm1$ being the
sign of the momentum, needed for the formal introduction of a time operator as
shown in Ref. 7. In this way Pauli's objection is resolved or circumvented.

\section{ The dynamical time operator and the discrete symmetries space
inversion, charge conjugation and time reversal \cite{Capri},\cite{Thaler}}

a)\qquad\textbf{Space inversion (Parity }$\boldsymbol{P}$\textbf{)}: denoting
by $<>_{P}$ the expectation value in the parity reversed state, one has:%
\begin{equation}
<\boldsymbol{r}>_{\boldsymbol{P}}=-<\boldsymbol{r}>;<\boldsymbol{p}%
>_{P}=-<\boldsymbol{p}>;<\boldsymbol{\alpha}>_{P}=-<\boldsymbol{\alpha
}>;<\beta>_{\boldsymbol{P}}=-<\beta>
\end{equation}

Thus $[T,\boldsymbol{P}]=0$ and%
\begin{equation}
<T>_{\boldsymbol{P}}=<T>.
\end{equation}

b) \qquad\textbf{Charge conjugation}: Under charge conjugation $\boldsymbol{C}%
$, one has%
\begin{equation}
<\boldsymbol{r}>_{\boldsymbol{C}}=<\boldsymbol{r}>;<\boldsymbol{p}%
>_{\boldsymbol{C}}=-<\boldsymbol{p}>;<\boldsymbol{\alpha}>_{C}%
=<\boldsymbol{\alpha}>;<\beta>_{\boldsymbol{C}}=-<\beta>(22)
\end{equation}

Then:%
\begin{equation}
<T>_{\boldsymbol{C}}=<\boldsymbol{\alpha\cdot r}/c>_{\boldsymbol{C}}%
+<\beta\tau_{0}>_{\boldsymbol{C}}=<T>-2<\beta>\tau_{0}.
\end{equation}

The expectation value in the charge conjugate state will only be equal to the
expectation value in the original state if $\tau_{0}$ is zero. Otherwise
$[T,\boldsymbol{C}]\neq0$.

c)\qquad\textbf{Time reversal}: Under time reversal \ $\boldsymbol{T}$\ one
has:%
\begin{equation}
<r>_{\boldsymbol{T}}=<r>;<p>_{\boldsymbol{T}}=-<p>;<\alpha>_{\boldsymbol{T}%
}=-<\alpha>;<\beta>_{\boldsymbol{T}}=<\beta>
\end{equation}

and therefore%
\begin{equation}
<T>_{\boldsymbol{T}}=-<T>+2<\beta>\tau_{0}.
\end{equation}

However it is seen that under the combined $\boldsymbol{C}$ and
$\boldsymbol{T}$ symmetries one has:%
\begin{equation}
<T>_{\boldsymbol{CT}}=-<\{\boldsymbol{\alpha\cdot r}/c+\beta\tau
_{0}\}>=-<T>,\qquad
\end{equation}

or, as parity leaves invariant the dynamic time operator,%
\begin{equation}
<T>_{\boldsymbol{CPT}}=-<T>
\end{equation}

The plausible expectation that the dynamical time operator would reverse sign
under time reversal occurs only if $\tau_{0}$ is zero. On the other hand, if
$\tau_{0}$ is different from zero, then charge conjugation is needed in
addition to produce the change in sign. This however brings it into agreement
with Feyman's proposal of the equivalence of the negative energy electron
states flowing backwards in time to positive energy positron states flowing
forward in time, $<H(e)>_{\boldsymbol{CPT}}=-<H(-e)>$ when charge is taken
into account. It is also in agreement with the positive energy extension of
Ref. 7, necessary for the introduction of a time operator, where the needed
degeneracy is provided by the $\boldsymbol{p}$ and $-\boldsymbol{p}$
degeneracy of the energy spectrum.

\section{ Zitterbewegung}

Dirac's equation yields a position vector $\boldsymbol{r}(t)$ consisting of a
term that follows the classical evolution to which is superimposed an
oscillatory motion, the Zitterbewegung (\textquotedblleft trembling
motion\textquotedblright). This motion is characterized in the low energy
range (see Appendix C) by an amplitude $\hbar/2m_{0}c$, the Compton wavelength
divided by $4\pi$ and a frequency $2m_{0}c^{2}/\hbar$, thus an oscillation
period $\hbar/2m_{0}c^{2}$. It is further demonstrated\cite{Capri}%
,\cite{Thaler} that this Zitterbewegung is not present in wave packets
constructed with purely positive (negative) energy states.
Alternatively\cite{Huang},\cite{Feschbach} it can also be shown that no finite
space width wave packet of positive (negative) mean energy can be constructed
without participation of negative (positive) energy states. Indeed, the
narrowest packet that can be built of positive energy states alone has a width
of the order $\hbar/m_{0}c$. Attempt to confine the packet within the spatial
range $\hbar/2m_{0}c$ makes this participation considerable (to construct a
$\delta$ function, negative and positive energy states must contribute with
equal weight), this being interpreted as the onset of particle antiparticle
pair creation. A similar situation arises with the system time operator. Its
spectrum spans all positive and negative $\tau$ values except for a gap from
$\tau_{0}$ to $-\tau_{0}$. In this representation a wave packet of finite
width with mean positive system time cannot be constructed without
participation of negative system time states, and cannot be confined within a
time span $2\tau_{0}$ without a considerable participation of these, that is,
without the creation of particle antiparticle pairs. This leads to identify
$\tau_{0}$ with the Zitterbewegung period $\hbar/2m_{0}c^{2}$ . A unified
\textit{spacetime \textquotedblleft Compton scale\textquotedblright}%
\ $\hbar/2m_{0}c$ and $\hbar/2m_{0}c^{2}$ is thus established, that sets
confinement limits in space and system time below which pair production
becomes significantly present\cite{Sidarth}.

Zittervewegung occurs \textquotedblleft naturally\textquotedblright\ in this
formulation, as a result of the mixing of positive and negative energy
eigenstates of the Dirac Hamiltonian. Its interpretation in the equation of
motion $\boldsymbol{r}(t)$ is still subject to discussion. It is known that it
can be eliminated by a redefinition of the position operator. Such is the so
called Newton Wigner position operator, based on the Foldy-Wouthuysen
representation, whose time derivative is just $c^{2}p/H$ (the group velocity)
instead of $c$, however at the price of an acausal propagation of initially
localized particles. This is a common problem with all position operators
commuting with the sign of energy\cite{Thaler}.

\section{ Conclusions}

Consideration of the free particle Dirac Hamiltonian leads to define a
particular dynamical self adjoint \textquotedblleft system time
operator\textquotedblright\ (i.e. based on a dynamic observable, namely the
position), and dependent on the parameter $t$ . It is shown to satisfy a
commutation relation with the Hamiltonian analogous to the one postulated for
the position and momentum operators in the sense of containing a constant part
independent of the particle state. The corresponding time energy uncertainty
relation -- now formally dependent on the position-momentum uncertainty
relation - involves simultaneous definite time expectation values as envisaged
originally by Bohr, i.e., the relation between the uncertainty in the instant
of time when the wave packet passes a particular spatial position with the
also instantaneous energy uncertainty associated with the wave packet. The
daring de Broglie's hypothesis\cite{Broglie} of a particle associated wave
whose phase velocity is larger than c is also derived as the instantaneous
rate of change of the position expectation value with respect to the
simultaneous expectation value of the time operator. Thus, these two elements
of the original basis and interpretation of quantum mechanics are integrated
into its formal mathematical structure.

The eigenvalue spectrum of the time operator contains a gap between positive
and negative values, similar to the gap occurring in the energy spectrum. Its
presence is needed to insure that charge conjugation has to be implemented in
addition of time reversal to connect positive and negative time values, in
agreement with Feyman's interpretation of the negative energy states.
Associating this gap to the period of the Zitterbewgung allows setting a
unified spacetime \textquotedblleft Compton scale\textquotedblright\ that
limits the width in space and time of the corresponding wave packets before
the generation of particle antiparticle pairs occurs.

The introduction of a dynamical system time operator does not question nor
invalidate the presence of the time parameter in the evolution postulate of
quantum mechanics, whose validity has been justified extensively in
experiments and applications (and whose presence may be explained by the
interaction with the environment, as quoted in the Introduction). On the other
hand, it may have relevance to current areas of research, such as:

i) Recently\cite{Horwitz} it has been argued that the interpretation of a
single particle experiment double slit interference in time\cite{Lindner}
(investigation that \textquotedblleft makes possible interferometry on the
attosecond time scale\textquotedblright) cannot be given in the non
relativistic Schr\"{o}dinger equation framework, as it requires
\textquotedblleft a wave function $\Psi(x,t)=<x|\Psi(t)>$ where $x$ and $t$
are the spectra of self-adjoint operators, to provide the possibility of
coherence in time, and therefore, interference phenomena\textquotedblright.
This assertion is based correctly on the fact that the evolution of the state
vector is given by a (external) parameter $t$ and not by the eigenvalue of a
self-adjoint operator canonically conjugate to the Hamiltonian, subject to the
well known objections\cite{Pauli}.

In the present case, the role of the (\textquotedblleft
external\textquotedblright\cite{Briggs1},\cite{Briggs2}) evolution parameter
$t$ (Schr\"{o}dinger picture) is maintained but an additional
\textquotedblleft observable\textquotedblright\ represented by a self-adjoint
operator $T$ is introduced, with eigenvectors $\left\vert \tau\right\rangle
$). Its spin-like eigenvector spectrum allows for the construction of an
extended Hilbert space $(\left\vert \boldsymbol{r}\right\rangle \otimes
\left\vert \tau\right\rangle )$. As it commutes with the position operator, a
\textquotedblleft four dimensional\textquotedblright\ representation of the
state vector $\left\vert \Psi(t)\right\rangle $ follows, namely $\left\langle
\boldsymbol{r},\tau\right.  \left\vert \Psi(t)\right\rangle =\Psi
(\boldsymbol{r},\tau;t)$. (Note that the free particle system is similarly
represented by $\Phi(\boldsymbol{p},E;t)=\left\langle \boldsymbol{p},E\right.
\left\vert \Psi(t)\right\rangle $ in the energy momentum space). Furthermore,
as the eigenvectors of the time operator are of the form$\left\vert
\tau\right\rangle =u_{r}\left\vert \boldsymbol{r}\right\rangle $, one has
that:%
\begin{equation}
\left\langle \boldsymbol{r},\tau\right.  \left\vert \Psi(t)\right\rangle
=u_{r}\left\langle \boldsymbol{r}\right.  \left\vert \Psi(t)\right\rangle
=u_{r}^{\dag}\psi(\boldsymbol{r};t)
\end{equation}

where $\psi(\boldsymbol{r};t)$ satisfies the time dependent Schr\"{o}dinger
equation. It then follows that%
\begin{equation}
\left\vert \left\langle \boldsymbol{r},\tau\right.  \left\vert \Psi
(t)\right\rangle \right\vert ^{2}\approx N^{2}(\tau_{r})\left\vert
\psi(\boldsymbol{r};t)\right\vert ^{2}%
\end{equation}

where $N(\tau_{r})$ is the normalization coefficient of the spinor $u_{r}$
(Appendix A). Finally, as shown numerically in Ref. 21, $\left\vert
\psi(\boldsymbol{r};t)\right\vert ^{2}$ does exhibit a time interference
pattern following a time double slit initial boundary condition.

ii) There is a current interest in the possibility of detecting Zitterbewegung
like effects in spintronics, grapheme and superconducting systems, due to the
similarity of their effective Hamiltonians with the Dirac Hamiltonian and the
fact that their space time conditions are close to current experimental
possibilities, these being in the spatial range of a few $\mathring{A}$\ and
of $1fs$ time pulses\cite{Sidarth},\cite{Cserti},\cite{Winkler},\cite{Zhang}%
,\cite{Rusin}. The possibility of defining an associated dynamical time
operator and its relation to expected experimental observations may be
addressed, based on the particular position operator $\boldsymbol{r}(t)$
appropiate in each case, as given by Eq. (4) in Ref. 17.

On the other hand, for electrons and heavier fermions the confinement limits
associated with the Compton scale are far beyond the above experimental
possibilities and no direct observation of Zitterbewegung can be expected. In
the case of electron neutrinos or antineutrinos with masses of the order of
2.2 eV, $\hbar/2m_{0}c=448\mathring{A}$\ and $\hbar/2m_{0}c^{2}=0.15fs$.
However, as they are found usually in an ultra relativistic range, the
applicable characteristic amplitude and period are much reduced as they are
attached to the de Broglie wave length (Appendix C).

iii) The dynamical system time operator may be helpful in resolving the so
called time paradox in quantum gravity that concerns the incompatibility of
the concepts of time in quantum mechanics -- where \textquotedblleft time
continues to be treated as a background parameter\textquotedblright\ - and in
general relativity -- where \textquotedblleft time is dynamical and
local\textquotedblright\cite{Macias}.

The dynamical time operator here proposed, commutes with the position
operator. However this does not lead to extend the normalization condition to
the additional variable, as occurs when going from one to three space
dimensions. Indeed, as pointed out in Ref.7, \textquotedblleft a consistent
definition of a probability density can include only points on a space-like
surface, i.e., with no possible causal connection. In the non-relativistic
limit $(c=\infty)$ all such surfaces are reduced to $\tau=const$ planes, and
the normalization applies only to the domain of space dimensions. Thus under
no circumstances is the time variable on a complete equal footing as the space
variables.\textquotedblright\

On the other hand the dynamical time operator, as defined, has a one to one
correspondence with the timelike worldline $\boldsymbol{r}(t)$ and is
monotonically linked to the time parameter $t.$ Then to each point of the
spectrum one can associate a spacelike surface that intersects the worldline
at the corresponding point, thus providing a foliation of spacetime by
spacelike surfaces over which one can define probability amplitudes.
Consequently one can say that this operator yields an observable variable that
\textquotedblleft sets the conditions\textquotedblright\ for the other
variables and defines a satisfactory notion of time, as required by the
conditional probability interpretation of quantum gravity\cite{Unruh}.

\section{Appendix A. Eigenvalues and eigenvectors of the dyamical time
operator}

Consider the eigenvalue equation of the self adjoint system time operator
$T=\boldsymbol{\alpha\cdot r}/c+\beta\tau_{0}$:%
\begin{equation}
T|\tau>=\tau|\tau> \tag{A.1}%
\end{equation}

In complete analogy with the energy eigenvalue and eigenvector solution in the
free particle case\cite{Dirac},\cite{Pauli},\cite{Capri},\cite{Thaler} one
has:%
\begin{equation}
|\tau>=u_{r}|\boldsymbol{r}> \tag{A.2}%
\end{equation}

where $|\boldsymbol{r}>$ is the eigenvector of the position operator
$\boldsymbol{r}$ and $u_{r}$ is a four component spinor independent of the
linear momentum $\boldsymbol{p}$. In the momentum representation the
eigenfunction is%
\begin{equation}
<\boldsymbol{p}|\tau>=u_{r}<\boldsymbol{p}|\boldsymbol{r}>=u_{r}(2\pi
\hbar)^{-3/2}expi(\boldsymbol{p\cdot r})/\hbar\tag{A.3}%
\end{equation}

Since from eq. (A.1) one has:%
\begin{equation}
T^{2}|\tau>)=\tau^{2}|\tau> \tag{A.4}%
\end{equation}

and%
\begin{equation}
T^{2}=\{\boldsymbol{\alpha\cdot r}/c+\beta\tau_{0}\}^{2}=(r/c)^{2}+\tau
_{0}^{2} \tag{A.5}%
\end{equation}

there are two (infinitely degenerate in the possible directions of
$\boldsymbol{r}$) eigenvalues of the time operator, namely:%
\begin{equation}
\tau=\pm\ \tau_{r}=\pm\ [(r/c)^{2}+\tau_{0}^{2}]^{1/2} \tag{A.6}%
\end{equation}

Each of these eigenvalues is doubly degenerate with respect to the component
$\boldsymbol{\sigma}$$\boldsymbol{\cdot r}/2r$ of the spin along the
$\boldsymbol{r}$ direction which commutes with $T$. Thus one can find
simultaneous eigenfunctions of $\boldsymbol{\sigma}$$\boldsymbol{\cdot r}/2r$
and $T$, giving rise to altogether four eigenvalue pairs:%
\[
(+\tau_{r},+1/2);(+\tau_{r},-1/2);(-\tau_{r},+1/2);(-\tau_{r},-1/2)
\]

The four orthonormal spinors $u_{r}$ are:

$\ \ \ \ \ \ \ \ \ \ \ \ \ \ \ \ \
\begin{tabular}
[t]{ccccc}%
$Sys.time$ & $Positive$ & $Positive$ & $Negative$ & $Negative$\\
$\tau$ & $+\tau_{r}$ & $+\tau_{r}$ & $-\tau_{r}$ & $-\tau_{r}$\\
$Spin$ & $+1/2$ & $-1/2$ & $+1/2$ & $-1/2$\\
$u_{1}$ & $1$ & $0$ & $-r/d$ & $0$\\
$u_{2}$ & $0$ & $1$ & $0$ & $+r/d$\\
$u_{3}$ & $+r/d$ & $0$ & $1$ & $0$\\
$u_{4}$ & $0$ & $-r/d$ & $0$ & $1$%
\end{tabular}
\ $

\bigskip

with $d=[c(\tau_{r}+\tau_{0})]$ and normalization coefficient $\ N(\tau
_{r})=[2\tau_{r}/(\tau_{r}+\tau_{0})]^{-1/2}$ for normalization to unity. For
a Lorentz covariant normalization, the normalization constant is

$[\tau_{r}/\tau_{0}]^{1/2}N(\tau_{r})=[1+(\tau_{r}/\tau_{0})]^{1/2}.$

The term $\tau_{0}$ is introduced solely by analogy with the rest mass term in
the free particle Dirac Hamiltonian. As such it gives rise to $2\tau_{0}$ gap
in the eigenvalue spectrum, separating positive and negative values in the
same way as the $2m_{0}c^{2}$ gap in the energy spectrum. No interpretation as
a property in analogy to the rest mass can be given at this stage, except
perhaps by recalling that the starting hypothesis of de Broglie's thesis was
to associate an oscillatory phenomenon of frequency $\nu_{0}=m_{0}c^{2}/\hbar$
with the rest mass of the particle, measured in the rest frame of reference.
The corresponding period would be $\hbar/m_{0}c^{2}$, that could be
interpreted as a characteristic internal \textquotedblleft system
time\textquotedblright\ $\tau_{0}$. This value is also related to the period
of the Zittervewegung. Note that, in this formulation, $\tau_{0}$ plays the
role of an invariant quantity in the $(\boldsymbol{r},\tau)$ space, i.e.,
$\tau_{0}^{2}=\tau^{2}-(\boldsymbol{r}/c)^{2}$, as $m_{0}c^{2}$ plays in the
$(\boldsymbol{p},E)$ space, namely $(m_{0}c^{2})^{2}=E^{2}-(c\boldsymbol{p}%
)^{2}$.

\section{Appendix B. The full dynamical time operator}

Eqs. (8), (9) and (10) for the operators $\boldsymbol{\alpha}$, $\beta$ and
$\boldsymbol{r}$ in the Heisenberg representation, yield for the time operator
$T(t)$ the following expression:%
\begin{align}
T(t)  &  =\boldsymbol{\alpha}(t)\cdot\boldsymbol{r}(t)/c+\beta(t)\tau
_{0}=\nonumber\\
&  =(1/c)(c\boldsymbol{p}/H)\cdot\boldsymbol{r}(0)+(c\boldsymbol{p}%
/H)^{2}t+(m_{0}c^{2}/H)\tau_{0}\nonumber\\
&  +\boldsymbol{[}exp(-2iHt/\hbar)\boldsymbol{]\{\alpha}(0)-c\boldsymbol{p}%
/H\boldsymbol{\}}\cdot\nonumber\\
&  \boldsymbol{\{r}(0)/c+(c\boldsymbol{p}/H)t+(\hbar/H)sin(-Ht/\hbar
)[(c\boldsymbol{p}/H)exp(iHt/\hbar)\nonumber\\
&  +(\boldsymbol{\alpha}(0)-c\boldsymbol{p}/H)exp(-iHt/\hbar)]\boldsymbol{\}}%
+\tau_{0}\boldsymbol{\{}\beta(0)-m_{0}c^{2}/H\boldsymbol{\}}exp(-2iHt/\hbar)
\tag{B.1}%
\end{align}

For $t=0$, this expression reduces correctly to $T(0)=\boldsymbol{\alpha
}(0)\cdot\boldsymbol{r}(0)/c+\beta(0)\tau_{0}$ .

\section{Appendix C - Non- and ultra- relativistic limits}

Setting $\boldsymbol{r}(0)=0$ in Eq.(10) for simplicity, the non and ultra
relativistic limits of the non oscillatory part of $T(t)$ are as follows:

1) Non relativistic limit $cp\ll m_{0}c^{2}$%
\begin{equation}
T(t)\simeq\tau_{0}+(cp/m_{0}c^{2})^{2}t+...)\qquad\tag{C.1}%
\end{equation}

Then, a dwell time between two points of the trajectory is given by:%
\begin{equation}
\delta T=T(t_{2})-T(t_{1})\simeq(cp/m_{0}c^{2})^{2}(t_{2}-t_{1}).\qquad
\tag{C.2}%
\end{equation}

2) Ultra relativistic limit $cp\gg m_{0}c^{2}$%
\begin{equation}
T(t)\simeq t+(m_{0}c^{2}/pc)\tau_{0}+....\qquad(C.3) \tag{C.3}%
\end{equation}

In this case, the dynamic time approaches the parametric (external) time $t$
and:%
\begin{equation}
\delta T=T(t_{2})-T(t_{1})\simeq t_{2}-t_{1} \tag{C.4}%
\end{equation}

As for the Zitterbewegung, whose amplitude and period are given by $c\hbar/H$
and $H/h$ respectively, the non relativistic limit yields $\lambda C/2\pi$ and
$\lambda C/c$, where $\lambda C$ is the Compton wave length $\ h/m_{0}c$,
establishing a \textit{spacetime \textquotedblleft Compton
scale\textquotedblright}. This amplitude restricts the localization of the
particle in space to one half Compton wavelength. In a similar way, the period
restricts the localization in time, in this case, to $\hbar/2m_{0}c^{2}$,
suggesting the value $\hbar/2m_{0}c^{2}$ for the parameter $\tau_{0}$, in
direct relation to the rest mass. On the other hand, in the ultra relativistic
limit the amplitude is $(1/2\pi)\lambda B$ where $\lambda B$ is the de Broglie
wave length $h/p$ , and the period is $\lambda B/c$, as noted in Ref. 18.

\section{Appendix D: Charged particle in an external electromagnetic field}

The \textquotedblleft minimal coupling\textquotedblright\ Dirac Hamiltonian
for a particle of charge $q$ in an external electromagnetic field is:%
\begin{equation}
H=\boldsymbol{\alpha\cdot\pi}(\boldsymbol{r},t)+\beta m_{0}c^{2}%
+q\Phi(\boldsymbol{r},t),\qquad\tag{D.1}%
\end{equation}

where $\boldsymbol{\pi}$$(\boldsymbol{r},t)=[\boldsymbol{p}%
-(q/c)\boldsymbol{A}(\boldsymbol{r},t)]$, and $\boldsymbol{A}(\boldsymbol{r}%
,t)$ and $\Phi(\boldsymbol{r},t)$ are the vector and scalar electromagnetic
potentials, respectively. Now:%
\begin{align}
\lbrack\boldsymbol{\alpha\cdot r},\boldsymbol{\alpha\cdot\pi}]  &
=\boldsymbol{r\cdot\pi}+i\boldsymbol{\Sigma}\cdot(\boldsymbol{r\times\ \pi
})-\boldsymbol{\pi\cdot r}-i\boldsymbol{\Sigma}\cdot(\boldsymbol{\pi\times
\ r})\nonumber\\
&  =3I+(4/\hbar^{2})\boldsymbol{s\cdot r\times}\ [\boldsymbol{p}%
-(q/c)\boldsymbol{A}(\boldsymbol{r},t)]. \tag{D.2}%
\end{align}

As\qquad$\lbrack\boldsymbol{\alpha\cdot r},\Phi(\boldsymbol{r},t)]=[\beta
,\Phi(\boldsymbol{r},t)]=0$,%
\begin{equation}
\lbrack T,H]=i\hbar\left[
\begin{array}
[c]{c}%
I+2\beta K-(4/\hbar^{2})(q/c)\boldsymbol{s\cdot r\times\ A(r,}t\boldsymbol{)}%
\\
+2\beta\left\{  \tau_{0}(c\boldsymbol{\alpha\cdot p})-m_{0}c^{2}%
(\boldsymbol{\alpha\cdot r}/c)\right\}
\end{array}
\right]  . \tag{D.3}%
\end{equation}

Finally:%
\begin{align}
d/dt  &  <T>=(1/i\hbar)<[T,H]>=1+2<\beta K>\nonumber\\
-(4/\hbar^{2})(q/c)  &  <\boldsymbol{s\cdot r\times\ A(r,}t\boldsymbol{)}%
>-i(2/\hbar)<\beta\{\tau_{0}(H-\Phi(\boldsymbol{r},t))-m_{0}c^{2}T> \tag{D.4}%
\end{align}

The time rate of change of the expectation value of the time operator is
modified by the electromagnetic interaction, in particular by the expectation
value of the vector potential. In consequence, different trajectories through
a non uniform electromagnetic field give rise to different time development of
the expectation value of the system time operator and different associated
phase velocities.

\begin{acknowledgement}
The author wishes to acknowledge the interest and enlightening critical
comments of Profs. P.A. Mello and L.P. Horwitz and of two unknowm reviewers.
\end{acknowledgement}

\bigskip

a. This article is a revised and extended version of the unpublished paper
arXiv:0908.2789v2 [quan-ph]

b. \ quoted by J. Polchinsky of UCSB in "23d Solvay Conference --The Quantum
Structure of Space and Time\textquotedblright, World Scientific, 2007.

\bigskip

\end{document}